\documentclass[aps,showpacs,preprint]{revtex4}
\usepackage{amsmath}
\usepackage{epsfig}

\begin{document}
\title{Longitudinal Response Functions of $^3$H and $^3$He }

\author{Victor D. Efros$^{1,3}$,
  Winfried Leidemann$^{1}$, 
  Giuseppina Orlandini$^{1}$ 
  and Edward L. Tomusiak$^{2}$}
\affiliation{$^{1}$Dipartimento di Fisica, Universit\`a di Trento, and
  Istituto Nazionale di Fisica Nucleare, Gruppo Collegato di Trento,
  I-38050 Povo, Italy\\
  $^{2}$Department of Physics and Astronomy
  University of Victoria, Victoria, BC V8P 1A1, Canada\\
  $^3$permanent address: Russian Research Centre "Kurchatov Institute", 
  123182 Moscow,  Russia\\
}

\date{\today}

\begin{abstract}
Trinucleon longitudinal response functions $R_L(q,\omega)$ are calculated for 
$q$ values up to 500 MeV/c. These are the first calculations beyond the 
threshold region in which both three-nucleon (3N) and Coulomb forces are fully 
included. We employ two realistic NN potentials (configuration space BonnA, 
AV18) and two 3N potentials (UrbanaIX, Tucson-Melborne). Complete final state 
interactions are taken into account via the Lorentz integral transform 
technique. We study relativistic corrections arising from first order
corrections to the nuclear charge operator.  In addition the reference frame
dependence due to our non-relativistic framework is investigated.
For $q \le 350$ MeV/c we find a 3N force effect between 5 and 15 
\%, while the dependence on other theoretical ingredients is small. At $q \ge 
400$ MeV/c relativistic corrections to the charge operator and effects
of frame dependence, especially for large $\omega$, become more important . 
In comparison with experimental data there is  generally a rather good 
agreement. Exceptions are the responses at excitation energies close to 
threshold, where there exists a large discrepancy with experiment at higher 
$q$. Concerning the effect of 3N forces there are a few cases, in particular 
for the $R_L$ of $^3$He, where one finds a much improved agreement 
with experiment if 3N forces are included.  
\end{abstract}

\bigskip

\maketitle
\section{Introduction}

Inclusive electron scattering can provide detailed information on the 
transition charge and current densities in nuclei.   
In the one photon exchange approximation
the cross section for this process is given by \cite{def} 
\begin{eqnarray}
{d^2\sigma\over d\Omega\,d\omega}\ =\ \sigma_M\ \bigg[\ {q_\mu^4\over q^4}\,
R_L(q,\omega)\ +
\ \bigr(\ {q_\mu^2\over 2q^2}+\tan^2{\theta\over 2}\ \bigr)\,R_T(q,\omega)\bigg]
\end{eqnarray}
where $R_L$ and $R_T$ are the longitudinal and transverse response functions
respectively, $\omega$ is the electron energy loss, $q$ is the magnitude of the 
electron momentum transfer, $\theta$ is the electron scattering angle and 
$q_\mu^2 = q^2 - \omega^2$. Experimental data for both $R_L$ and $R_T$ are 
available for a variety of energy and momentum transfers.  However because of 
our non--relativistic treatment of the nuclear dynamics we restrict our 
attention to momentum transfers $q\leq$ 500 MeV/c and energy transfers 
$\omega\leq$ 300 MeV. Data covering various regions in this range are 
given for both $^3$H and $^3$He by Retzlaff {\it et al} \cite{retz}, Dow 
{\it et al} \cite{dow}, Marchand {\it et al} \cite{mar}, and Morgenstern 
\cite{morgen}. 

The theoretical treatment of these response functions requires the ability to 
accurately include transitions to the continuum. Techniques for doing this 
with realistic NN potentials have only been developed and implemented during 
the past ten years. These include both Faddeev and Lorentz integral transform 
(LIT) methods \cite{ELO1,martinelli,golak95,glock98,leid_bled}.  For the 3N 
photodisintegration total cross sections results obtained with the 
LIT \cite{ELOT01} and Faddeev techniques are compared in \cite{GG}. 
In the work of 
Viviani {\it et al} \cite{viv} expansion techniques were applied to solving the 
ground state and continuum wave equations, but the calculation was restricted 
to a $^3$He near threshold region where only the two--body breakup occurs.
Previous to the above references Faddeev calculations of trinucleon response 
functions were published by Meijgaard and Tjon \cite{tjon} in 1992 using the 
s-wave Malfliet--Tjon potential MT--I/III \cite{MTpot}. Apart from how the 
quantum mechanics is done there are major differences in physics input between 
the longitudinal and transverse responses. Whereas the non--relativistic 
longitudinal response requires only a charge operator and nucleon form factors, 
the transverse response requires exchange currents in addition to single 
nucleon currents and nucleon form factors.  It is clear that if a given nuclear
interaction cannot describe the longitudinal response then it would be 
pointless to attempt a calculation of the transverse reponse. In particular if 
one inquires into the effect of three--body forces in nuclei it would appear 
natural to first investigate their impact on the longitudinal response.  
Otherwise, through a calculation of $R_T$, it would be difficult to disentangle 
the effects of three--body forces from exchange current effects.  Further as 
shown in \cite{golak95}  the longitudinal response appears in general 
insensitive to the realistic NN force model thus removing a possible source of 
ambiguity when comparing the effects of different three body force models on 
$R_L(q,\omega)$.

\section{Nuclear forces and charge operator} 

 The function $R_L$ represents the response 
of the nucleus through the nuclear charge operator $\rho$ and is given by
\begin{equation}   
R_L(q,\omega)=\overline{\sum}_{M_0}\sum\!\!\!\!\!\!\!\!\int df\langle\Psi_0|\rho^\dag({\bf q},\omega)|\Psi_f\rangle
\langle \Psi_f|\rho({\bf q},\omega)|\Psi_0\rangle\,\,
\delta(E_f-E_0+q^2/(2M_T)-\omega).\label{rl}
\end{equation}    
Here $\Psi_0$ and $\Psi_f$ denote the ground and final states, 
respectively, while $E_0$ and $E_f$ are energies pertaining to them,  
\begin{equation}
(H-E_0)\Psi_0=0,\qquad (H-E_f)\Psi_f=0,
\end{equation}
where $H$ is the nuclear non--relativistic Hamiltonian. The above quantities 
$\Psi_{0,f}$ and $H$ are internal quantities in the hadronic c.m. frame. The 
integration (summation) goes over all final states belonging to the same 
energy $E_f$, and $M_0$ is the projection of the ground state angular momentum.

The Hamiltonian includes the kinetic energy terms, the NN and 3N force terms, 
and the proton Coulomb interaction term in the $^3$He case. The ground state 
$\Psi_0$ is calculated via an expansion in basis functions which are correlated 
sums of products of hyperradial functions, hyperspherical harmonics and 
spin--isospin functions. In the present 
work three models of the NN force are used, the realistic AV18 \cite{AV18} and 
configuration space BonnA (herein referred to as BonnRA) \cite{BonnRA} models, 
and the s-wave MT--I/III potential. We consider two 3N force models, 
the UrbIX \cite{urb9} and 
the TM$^\prime$ \cite{TM}, in the combinations AV18+UrbIX, AV18+TM$^\prime$ 
($\Lambda$=3.358 fm$^{-1}$), and BonnRA+TM$^\prime$ ($\Lambda$=2.835 fm$^{-1}$).
As indicated the TM$^\prime$ cut-off parameter $\Lambda$ is different in the 
AV18 and BonnRA combinations in order to properly fix the $^3$H binding energy 
in each case. Table I lists our results for ground state properties for the 
above potential combinations containing the 3N force. 

As nuclear charge operator we take the following one--body operator
\begin{equation}
\rho({\bf q},\omega)=\sum_{j=1}^A\rho_{j}^{nr}({\bf q},\omega)+\rho_{j}^{rc}
({\bf q},\omega),\label{rho}
\end{equation}
where 
\begin{eqnarray}
 \rho_{j}^{nr}({\bf q},\omega)=\hat{e}_je^{i{\bf q}\cdot{\bf r}_j},\label{nr}
\end{eqnarray}
\begin{eqnarray}
 \rho_{j}^{rc}({\bf q},\omega)=-\frac{q^2}{8M^2}\hat{e}_je^{i{\bf q}\cdot
{\bf r}_j} -i\frac{\hat{e}_j-\hat{\mu}_j}{4M^2}\,{\vec\sigma}_j\cdot({\bf q}
\times{\bf p}_j) e^{i{\bf q}\cdot{\bf r}_j},\label{rc}
\end{eqnarray}
\begin{eqnarray}
\hat{e}_j=G_E^p(q_\mu^2)\frac{1+\tau_{z,j}}{2}+G_E^n(q_\mu^2)
\frac{1-\tau_{z,j}}{2}
\equiv\frac{1}{2}\left[G_E^S(q_\mu^2)+G_E^V(q_\mu^2)\tau_{z,j}\right],\label{e}
\end{eqnarray}
\begin{eqnarray}
\hat{\mu}_j=G_M^p(q_\mu^2)\frac{1+\tau_{z,j}}{2}+G_M^n(q_\mu^2)
\frac{1-\tau_{z,j}}{2}.\label{m}
\end{eqnarray}
Here $\bf r$, $\bf p$, $\vec\sigma$, and $\vec\tau$ are the nucleon position, 
momentum, spin and isospin operators, $M$ is the nucleon mass, and 
$G_{E,M}^{p,n}$ are the nucleon Sachs form factors.
The two terms in 
(\ref{rc}) proportional to $M^{-2}$ are the Darwin--Foldy (DF) and spin--orbit 
(SO) relativistic corrections to the main operator (\ref{nr}), see e.g. 
\cite{def,friar}. We refer to the main operator (\ref{nr}) as the 
non--relativistic one although the dependence of nucleon form factors on 
$q_\mu^2$ does not allow a non--relativistic interpretation. 

In this work we mainly  use the well known dipole fit for proton electric 
form factor, while the neutron electric form factor is taken from 
\cite{Galster}, but we also check the $R_L$ dependence with a different 
parametrization, namely the best fit from \cite{Hoehler} to these form 
factors. In case of the SO term we adopt the usual although recently 
controversial \cite{contro} approximation 
$G_M^{p,n}(q_\mu^2)=\mu_{p,n}G_E^{p}(q_\mu^2)$ in (\ref{m}), $\mu_{p,n}$ being 
proton and neutron magnetic momenta. For the calculation of $R_L$ it is
convenient to rewrite the operator $\rho$  in terms of   
the isoscalar and isovector charge nucleon form factors from (\ref{e}), 
\begin{eqnarray}
\rho({\bf q},\omega)= G_E^S(q_\mu^2)\rho_s({\bf q}) + 
G_E^V(q_\mu^2)\rho_v({\bf q}).\label{rsv}
\end{eqnarray} 
 
The inclusion of relativistic corrections for the one--body charge operator only
is not completely consistent. In fact there exist additional
relativistic effects: a wave function boost (as done in \cite{Beck,Beck1} for 
the $d(e,e')$ reaction) and additional two-body terms in the charge operator 
(as done in \cite{viv} for the low-energy two-body break-up channel of the 
$^3$He$(e,e')$ reaction). In our case there are two reasons why we include 
the relativistic corrections to the one-body charge operator:

\noindent
(i) At higher $q$ they lead to an important reduction of the $R_L$ quasi-elastic 
peak height. As illustrated in \cite{Beck1} such a reduction is confirmed if 
boost corrections are included. Moreover, the frame 
dependence of the response functions is studied in \cite{Beck} where it is shown
that in the 
Breit frame boost corrections are negligibly small for the quasi-elastic peak
region (different kinematics are not shown). We believe that one has a similar 
frame dependence of boost corrections also for the electromagnetic response 
of the three-nucleon systems. Thus we will make the comparison with 
experimental data taking $R_L$ from a Breit frame calculation with a subsequent
transformation into the $R_L$ LAB frame result (see discussion of Fig.~5).

\noindent
(ii) They enable us to make a direct comparison of our results with those of
\cite{viv}. Since realistic few--body calculations are rather complicated it
is of great importance to have these kind of checks.

\section{Calculation of response}

We calculate $R_L$($q,\omega$) by the LIT method as described in 
\cite{ELO1,ELO2}. The technique is, however, directly applicable only when the 
transition operator does not depend on $\omega$. To separate out the $\omega$ 
dependence of the transition operator we use Eq. (\ref{rsv}) to represent 
the response function (\ref{rl}) as
\begin{eqnarray}
R_L(q,\omega)=[G_E^S(q_\mu^2)]^2\,{\cal R}_{s}+
G_E^S(q_\mu^2) G_E^V(q_\mu^2)
{\cal R}_{sv}+
[G_E^V(q_\mu^2)]^2  {\cal R}_{v} \ ,
\label{sep}\end{eqnarray}
where  ${\cal R}_{s}$ and ${\cal R}_{v}$ are the responses which emerge if
$\rho_s({\bf q})$ and $\rho_v({\bf q})$ are taken as transition operators, and
the quantity ${\cal R}_{sv}$ is the mixed response,
\begin{eqnarray}   
{\cal R}_{sv}(q,\omega)=\overline{\sum}_{M_0}\sum\!\!\!\!\!\!\!\!\int df
\left[\langle\Psi_0|\rho_s^\dag({\bf q})|f\rangle
\langle f|\rho_v({\bf q})|\Psi_0\rangle+
\langle\Psi_0|\rho_v^\dag({\bf q})|f\rangle
\langle f|\rho_s({\bf q})|\Psi_0\rangle\right]\nonumber\\
\times\delta(E_f-E_0+q^2/(2AM)-\omega).
\label{mr}\end{eqnarray}  
To calculate the subsidiary responses entering (\ref{sep}) with the LIT method
one can solve the inhomogeneous equations
\begin{eqnarray}
( H - E_0 - \sigma )\tilde\Psi_s(\sigma)= \rho_s\Psi_0,\qquad
( H - E_0 - \sigma )\tilde\Psi_v(\sigma)= \rho_v\Psi_0\label{deq}\label{eqs}  
\end{eqnarray}   
for a set of complex $\sigma$ values and then form the scalar products 
$\langle\tilde\Psi_s(\sigma)\,|\,\tilde\Psi_s(\sigma)\rangle$,
$\langle\tilde\Psi_v(\sigma)\,|\,\tilde\Psi_v(\sigma)\rangle$, and
$\langle\tilde\Psi_s(\sigma)\,|\,\tilde\Psi_v(\sigma)\rangle$. 
These scalar  products represent integral transform with Lorentzian
kernels, e.g., one has    
\begin{equation}
\langle\tilde\Psi_s(\sigma)\,|\,\tilde\Psi_s(\sigma)\rangle =
{{\cal R}_{s}^{el}(q,\omega_{el}) \over |\sigma|^2} +
 \int_{\omega_{th}}^\infty d\omega {{\cal R}_{s}(q,\omega) 
\over (\omega-\omega_{el}-\sigma)(\omega-\omega_{el}-\sigma^*)} \,.
\end{equation}
Here ${\cal R}_{s}^{el}$ is the elastic scattering form factor, 
$\omega_{el}=q^2/(2M_T)$, and
$\omega_{th}$ is the threshold for the inelastic energy transfer. 
From the inversion of such integral transforms one then obtains
the response functions ${\cal R}_{s}$, ${\cal R}_{v}$ and ${\cal R}_{sv}$.

In previous work \cite{ELO2} equations similar to those in (\ref{eqs}) were  
solved numerically in order to calculate the above mentioned scalar products  
$\langle\tilde\Psi_i(\sigma)\,|\,\tilde\Psi_j(\sigma)\rangle$.
An alternative and computationally more efficient way of the calculation of the 
transform is a direct evaluation of the scalar products via the Lanczos 
technique \cite{Lanczos}. Thus we use this method in our calculation.

As for $\Psi_0$ we perform expansions in terms of basis functions $|\mu\rangle$ 
which are correlated sums of products of hyperradial functions, hyperspherical 
harmonics and spin--isospin functions. The first Lanczos vector is given by 
\begin{equation}
\label{first}
|\varphi_0\rangle={\frac{|{\Psi}\rangle}
{\sqrt{\langle \Psi | \Psi\rangle}}
} 
\end{equation}
\noindent
with
\begin{equation}
|{\Psi}\rangle = g_{\mu\nu}^{-1} \, \hat{\rho} \, |{\Psi_{0}}\rangle \,,
\end{equation}
where $g_{\mu\nu}^{-1}$ denotes the inverse of the norm matrix $g_{\mu\nu}=
\langle \mu | \nu \rangle$. 
One then applies recursively the following relations  
\begin{equation}
\label{6'}
{b_{n+1}|\varphi_{n+1}\rangle = g_{\mu\nu}^{-1} H |\varphi_{n}\rangle 
+ a_{n}|\varphi_{n} \rangle -b_{n} |\varphi_{n-1}\rangle}\,,
\end{equation}
\begin{equation} 
a_{n}=\langle \varphi_{n}|H| \varphi_{n}\rangle,~~ 
b_{n} \equiv ||b_{n}|\varphi_{n}\rangle||,
\end{equation}
\noindent 
where $a_{n}$ and $b_{n}$ are the Lanczos coefficients. The transform 
can then be written as a continuous fraction

\begin{equation}
\label{7}
{\langle\tilde\Psi\,|\,\tilde\Psi\rangle=
\frac{2i}{\sigma-\sigma^*}Im\frac{\langle \Psi | \Psi\rangle }{(z-a_{0})-
\frac{b^{2}_{1}}{(z-a_{1})-
\frac{b^{2}_{2}}{(z-a_{2})-b^{2}_{3}....}}}~}
\end{equation}
\noindent  
with $z=\sigma+E_0$.

Basis functions posess definite values of parity $P$, angular momenta $J$ and  
magnetic quantum numbers $M_J$. Various Lanczos sets are separated  with 
respect to 
these quantum numbers. Multipole expansions of the operators $\rho_s$ and  
$\rho_v$ are performed, which allows elimination of dependencies on 
$M_J$ and on the ground--state angular momentum projection $M_0$. 
In our case there exists only one multipole $\ell$ 
compatible with a given $J$ and $P$ value. Indeed, one has $J=\ell\pm 1/2$, 
and parity equal to $(-1)^\ell$. 
 
Hyperspherical harmonics (HH) belonging to given permutational 
symmetry types are obtained via application of the corresponding 
symmetrization  operators to HH of the type $Y_{KLM_{L}}^{l_1l_2}$ where $K$ 
is the grand--angular momentum, $L$ and $M_L$ are the total orbital momentum 
and its projection, and $l_1$ and $l_2$ are the orbital momenta associated with
a relative motion of a given pair of particles and a relative motion of the 
third particle with respect to the pair. These basis HH are  coupled to 
the spin--isospin functions of conjugated permutational symmetry types to get
basis functions antisymmetric with respect to permutations of nucleons.
These spin--isospin functions posess given spin $S$ and isospin $T$.
Thus our basis functions have  given $J$, $M_J$, $K$, $L$, $S$, $T$ 
values, given parity equal to $(-1)^K$, and given type of symmetry with respect 
to permutations of spatial, or spin--isospin, variables. In order to
accelerate the convergence of the HH expansion a spin and isospin dependent 
correlation operator is  applied to the basis functions (see \cite{ELOT01} for 
details). Matrix elements are  calculated analytically with respect to 
three Euler angles determining the orientation of the system as a whole, and the 
remaining three--dimensional integrations are  done numerically. 

Rather many basis functions are  retained to achieve convergence, and a 
selection of basis HH  has been  done to reduce their net numbers 
in the calculation. The selection is  based upon the property 
\cite{efr72} that the uncorrelated symmetrized basis HH obtained in the above 
mentioned way from the subset of HH  $Y_{KLM_{L}}^{l_1l_2}$ with only small  
$l_1$ and small $l_2$ suffice to provide a predominant contribution to bound state 
wave functions. We have found that in practice this property is  also 
valid  for our correlated HH and for the case of our inhomogeneous 
equations. At the same time, the selection  depended on  $L$, $K$, $J$ values
and symmetry types of HH  as well.

The mixed response
(\ref{mr}) is  calculated as   
\begin{equation}
{\cal R}_{sv}={\cal R}_{s+v} - {\cal R}_{s}
- {\cal R}_{v} \ ,
\end{equation}
where ${\cal R}_{s+v}$ is the response, which emerges from the scalar product
$\langle\tilde\Psi_{s+v}(\sigma)\,|\,\tilde\Psi_{s+v}(\sigma)\rangle$, and
$\tilde\Psi_{s+v}(\sigma)$ is obtained if $\rho_s+\rho_v$, instead of $\rho_s$ 
or $\rho_v$, is taken as the transition operator in (\ref{eqs}).

\section{Results and Discussion} 

Before showing detailed results of our calculations it is necessary to address 
the question of convergence  with respect to the maximum angular momentum $J_{max}$
retained in 
our calculation. This requires some measure of convergence. In this connection 
we consider  here the $^3$H Coulomb sum rule results computed for the case
$G_E^S = G_E^V=1$. The sum rule reads in this case 
\begin{eqnarray}
 \int_{\omega_{th}}^\infty\ R_L(q,\omega)\ d\omega +  
R_L^{el}(q,\omega_{el}) =1.
\label{sr}
\end{eqnarray}
Since in the case considered a single proton interacts with the 
electromagnetic field,   Eq. (\ref{sr}) does not contain the nucleon charge 
correlation contribution and is valid for any $q$. It is clear that larger $q$ 
values require the expansion to include larger values of $J$. Table II shows 
the results of using $J_{max}$ = 15/2 for the $q$= 250, and 300 MeV/c cases and 
$J_{max}$=21/2 for the $q$=350 - 500 MeV/c cases. One notes that the lower $q$ 
sum rules are nearly fully converged while the 500 MeV/c case still requires 
about 2\% more strength. Although this could be improved by increasing 
$J_{max}$ we consider the convergence tolerable for the present investigation. 
Table II also demonstrates that the convergence is faster for the simple 
MT--I/III potential as compared to the realistic potential models.

In Fig.~1 we illustrate the dependence of $R_L$ on the NN potential. The 
results with the two realistic potentials, Bonn and AV18, are very similar at
$q=500$ MeV/c, but exhibit somewhat stronger differences for the quasi--elastic 
peak height at $q=250$ MeV/c. With the semi--realistic MT--I/III potential one 
observes a rather similar picture for $q=250$ MeV/c as with the realistic 
potentials, whereas at $q=500$ MeV/c a greater peak height and considerably 
less high--energy strength than for the realistic potentials is found. 

In Fig.~2 we show the 3N force effect.  It is seen that it decreases the 
peak height and enhances the high--energy tail. At lower momentum transfer the 
reduction of the peak 
height is more pronounced. Comparing the three cases, where a  3N force is 
included, one finds only rather small differences among them except for the 
low--energy range at $q=500$ MeV/c as will be seen next in Fig.~3. 

In order to study the low--energy behavior better, in Fig.~3 we illustrate the 
nuclear force  model dependence of the triton $R_L$ close to threshold at 
three momentum transfers covered also by the data of \cite{retz}. In this 
figure $R_L$ is shown as a function of $E_x$, the relative kinetic energy of 
the outgoing three nucleons. At $q=174$ MeV/c there is a rather strong decrease 
of $R_L$ due to the 3N force. The reduction becomes considerably smaller at 
$q=324$ MeV/c and at $q=487$ MeV/c the 3N force leads to an opposite effect, 
namely a moderate increase. From the comparison of the cases AV18+UrbIX and 
AV18+TM$^{\prime}$ it becomes clear that the 3N force model dependence is for 
all the three momentum transfers very small. The only evident potential model 
dependence is found at the highest $q$, where the case BonnRA+TM$^{\prime}$ 
exhibits considerably more strength than the other cases with inclusion of 3N 
force. 

In Fig.~4 we show the effect of the relativistic corrections on $R_L$. One sees 
that the SO term leads only to rather small contributions, while the DF term is 
more important. It occurs that separate contributions from 
the SO term are not so small, only their net sum proves to be very small. This 
probably means that in the inclusive case we have an effect of averaging out 
due to the spin dependence of the SO operator. Because of the smallness of the 
SO contribution we have neglected it in most of the following cases. In Fig.~4 
we also show $R_L$ results, where a different nucleon form factor 
parametrization \cite{Hoehler} is taken. At $q=250$ MeV/c the different form 
factors lead to very similar results, but at $q=500$ MeV/c there is a 3 \% 
reduction of $R_L$ with the parametrization of \cite{Hoehler}. In the results 
which follow  we will always use the dipole nucleon form factors. However as seen 
here there will be uncertainties in $R_L$ at higher $q$ values due to uncertainties
in the nucleon form factors. 
 
Next we would like to check the frame dependence of our calculation.  
To this end we calculate $R_L$ also in the Breit (B) frame and the socalled 
anti--lab (AL) frame. 
In the AL frame the virtual photon and initial target nucleus have momenta 
${\bf q}_{AL}$ and $-{\bf q}_{AL}$, respectively, whereas the total momentum of 
the final three--nucleon state is equal to zero. Note that in the LAB frame one 
has the opposite case: the target nucleus in the initial state is at rest and 
the total momentum of the final three-nucleon state is equal to $\bf q$. 
Finally, in the Breit frame one has total momenta of initial and final hadron 
states equal to $-{\bf q}_B/2$ and ${\bf q}_B/2$, respectively, while the 
photon four-momentum is $(\omega_B,{\bf q}_B)$. Formally there are no 
differences between the calculations in the various frames. One obtains a 
response function which has the arguments $\omega$ and $q$ of the given frame, 
i.e. $R^{LAB}_L(q_{LAB},\omega_{LAB})$, $R^{AL}_L(q_{AL},\omega_{AL})$ and 
$R^B_L(q_B,\omega_B)$. For a comparison of the results we transform
$R^{AL}_L(q_{AL},\omega_{AL})$ and $R^B_L(q_B,\omega_B)$ into
$R^{LAB(AL)}_L(q_{LAB},\omega_{LAB})$ and $R^{LAB(B)}_L(q_{LAB},\omega_{LAB})$,
respectively. To this end we use that the various reference frames are 
connected via Lorentz boosts and thus $\omega_{AL}$, $q_{AL}$, $\omega_B$ and 
$q_B$ can be expressed through $\omega_{LAB}$ and $q_{LAB}$. However in order 
to obtain an $R_L$ in the LAB frame from $R_L$'s in  AL and Breit frames it is 
not sufficient to transform the relative arguments of $\omega$ and $q$ into the 
corresponding LAB frame arguments. In addition one has 
\begin{equation}
R^{LAB(frame)}_L = {q_{LAB}^2 \over q_{frame}^2} R^{frame}_L \,,
\label{frame} 
\end{equation} 
where "$frame$" stands for AL or Breit.
The origin of the additional factor is the following. The cross 
section of (1) contains three separate pieces, namely $\sigma_M$, a part 
regarding the electron (e.g., $q_\mu^4/q^4 \equiv V^{LAB}_L$) and a hadronic 
part (e.g., $R_L$). The latter two originate from a reduction of a product of 
leptonic and hadronic Lorentz tensors \cite{def}. The product of these two tensors
forms a Lorentz scalar and thus is frame independent. One can show that for the 
longitudinal part of the cross section of (1) one has 
\cite{Beck}
\begin{equation}
V^{LAB}_L = {q_{frame}^2 \over q_{LAB}^2} V^{frame}_L \,
\end{equation}  
and thus Lorentz invariance requires the additional factor in (\ref{frame}). 

In Fig.~5 we compare the longitudinal response functions of the various frames.
At $q$=250 MeV/c differences are rather small, in particular between Breit and 
AL frame results. Except for the threshold region there is not such 
a similarly good agreement at $q$=500 MeV/c. In the quasi-elastic peak there 
are rather pronounced differences: $R_L^{LAB(AL)}$ is about 7 \% and 
$R_L^{LAB(B)}$ about 4 \% higher than $R_L^{LAB}$, their peak positions are 
shifted by about 6 (AL) and 5 MeV (B) towards lower energies. In a consistent 
relativistic theory one would of course have identical results and thus the 
obtained differences point to a relativistic inconsistency in the calculation. 

As mentioned before Beck {\it et al} \cite{Beck} studied the electromagnetic
response functions in deuteron electrodisintegration in the quasi-elastic
region. They have shown that an inclusion of boost effects on the hadron wave 
functions leads essentially to the same results for the various reference 
frames discussed here. In addition they have found that boost corrections are 
almost vanishing in the Breit frame. We believe that also in the three-nucleon 
electrodisintegration one probably has a similar picture with a strong
cancellation of boost effects in the Breit frame. Therefore we will take
the $R_L^{LAB(B)}$ results in comparison with quasi--elastic experimental data.

A comparison of the $^3$H and $^3$He theoretical longitudinal response 
functions with experimental data of \cite{dow,mar,morgen} is shown in Fig.~6 at 
$q=250$, 300, and 350 MeV/c. In the peak region one does not find a clear 
picture, since there is a better agreement once with the 3N force ($^3$He) and 
once without the 3N force ($^3$H). Except for the triton case at $q=250$ MeV/c 
one observes rather similar theoretical and experimental results for the 
high--energy tail. At higher energies the size of the experimental errors is 
larger than the effect of the 3N force, thus nothing can be said there about 
an improvement of the theoretical result with the 3N force. 

In Fig.~7 we show equivalent results as in Fig.~6 but at the higher momentum 
transfers of 400, 450, and 500 MeV/c. Also here one finds a better agreement 
with experimental data without the 3N force in case of $^3$H and with the 3N 
force in case of $^3$He. It is worthwhile to note that for all six cases of 
Fig.~7 one has a good agreement of theoretical and experimental peak positions. 
Concerning the low-- and high--energy tails one has a rather good agreement 
between theory and experiment. 

Next we turn to a comparison of the triton low--energy longitudinal response 
functions with the experimental data of \cite{retz}. In Fig.~8 we show the 
$R_L$ of $^3$H at various $q$. Since the $R_L$ frame dependence is very 
small close to threshold we illustrate directly the results from a LAB frame 
calculation.  For the 
lower two momentum transfers there is a rather good agreement of experiment and 
theory, but the size of the experimental error is too large to draw definite 
conclusions about possible improvements due to the 3N force. At $q=487$ MeV/c 
the picture is different, the theoretical response functions are larger than 
the experimental one, in particular very close to threshold. It is also 
evident that the effect of the 3N force moves the calculated $R_L$ even further away
from the data. 

In Fig.~9 we show a similar comparison with experimental data as in Fig.~8 but
for the $R_L$ of $^3$He. Again one finds a rather good agreement between theory 
and experiment for the two lower $q$'s, but contrary to the triton case here 
the 3N force is important for this agreement at $q=174$ MeV/c. Also for the 
highest momentum transfer one finds a similar picture as for the triton case, 
namely a large overestimation of the experimental data by the theoretical 
response functions and also an increase of $R_L$ due to the 3N force. 

In Fig.~9 we also illustrate theoretical results from \cite{viv}. It is an
approach to calculating responses which is entirely different from ours. 
The calculation \cite{viv} has been carried out with the AV18+UrbIX potentials, 
relativistic 
DF- and SO-terms have been included and the same nucleon form factors as by us 
have been used (dipole fit, neutron electric form factor from \cite{Galster}). 
In order to have a clean comparison of the two different calculations, we also 
take into account the SO term for our result with AV18 and UrbIX, though its
effect is also here very small. For the two higher momentum transfers there
is a rather good agreement between both calculations. Some differences
are visible at $q=174$ MeV/c, but the difference between the two calculations 
is still considerably smaller than the experimental error bars.

The rather large discrepancy between theory and experiment of the 
low--energy $R_L$ at $q=487$ MeV/c requires further theoretical and experimental
investigations. We should mention that in the calculation of \cite{viv} 
relativistic two--body charge operators were also considered. Although
they were not sufficient 
to give agreement with experiment, they did diminish the discrepancy by about 
a factor of two. 
Concerning the nucleon form factors one could only  obtain a small 
reduction (about 3  \%) using the parametrization of \cite{Hoehler}. 
 In addition the potential model depencence should be further studied. In 
the discussion of Fig.~3 we have already mentioned a rather
strong potential model dependence of the low--energy $R_L$ at $q=487$ MeV/c.
Therefore it would be interesting to consider other modern 
realistic NN potentials in addition to the AV18 and BonnRA models used here.

\section{Conclusions}

In the following we give a brief summary of our work. The trinucleon 
longitudinal response function $R_L(q,\omega)$ is calculated with realistic NN 
interactions, 3N, and  Coulomb forces for a variety of kinematical settings 
that include momentum transfers $q$ between 174 and 500 MeV/c and wide ranges 
of energy transfers $\omega$. The results are fully convergent. The 
calculations are performed via the Lorentz Integral Transform method. 
 
As NN interaction we use a modern realistic (AV18), a realistic (BonnRA), and 
also a semi--realistic (MT--I/III) potential model. Two models (UrbIX, TM$^{\prime}$)
of the 3N force are employed. The treatment of the trinucleon 
dynamics is completely non--relativistic. Nonetheless we apply a minimal check 
on the uncertainties related to this. For this purpose we evaluate $R_L$ in 
three different reference frames, namely in lab, anti--lab, and Breit 
frames.  For the charge operator we take the leading relativistic 
corrections into account (Darwin--Foldy and spin--orbit term). 

In general we find a rather small NN potential model dependence, but in
some cases there are also larger effects. These include the height of the
quasi--elastic peak at lower $q$ and  the threshold behavior at higher $q$.
The effect of the 3N force is typically between 5 and 10 \%, but reaches 
up to 15 \%  for the  low--enery response at low $q$. The dependence
on the 3N force model is very small for all considered cases. 

Concerning the relativistic contributions to the charge operator, our 
inclusive case shows negligible effects due to the spin--orbit term, while 
the DF term leads to non--negligible effects at higher $q$. 
With respect to the $R_L$ calculation in the various reference frames, we 
observe 
a non--negligible frame dependence at higher $q$, except for the threshold 
region. In order to restore a more consistent relativistic behavior one would 
need to consider additional relativistic effects. Similar results have been 
found in $d(e,e')$ and it is shown that additional boost corrections lead to a 
much better agreement among the various frame results \cite{Beck}. In the same
work it is also shown that boost effects are negligible in the Breit frame. We 
assume a similar behavior also in trinucleon electrodisintegration. Thus we 
compare the $R_L$ calculated in the Breit frame with experimental data. 

The comparison of our results with experimental data is generally rather 
satisfying for all considered momentum transfers, in particular for the $R_L$
of $^3$He . The experimental data, however, are in most cases not precise 
enough to draw definite conclusions about the 3N force effect. A nice exception
is the $^3$He low--energy response, where a  3N force proves to be necesssary
to obtain agreement with experiment.  In addition for the $^3$He
quasi--elastic peak heights at $q \le 400$ MeV/c three-nucleon forces 
considerably improve the agreement with experiment. 
At higher $q$ and low $\omega$ values one finds a considerably higher 
$R_L$ response in theory than in experiment. 

Last but not least we would like to mention that at very low energies, i.e. up 
to the three--body breakup threshold, we can compare our results with those
of \cite{viv}. We find quite a good agreement.  The differences which do show
up a very low $q$ are still smaller than the experimental error bars. 

\section*{Acknowledgment}

We thank K. Dow for sending us easily plottable data from her experiment 
described in \cite{dow} and M. Viviani for giving us his theoretical results
\cite{viv}. 
One of us (W.L.) thanks H. Arenh\"ovel for
a discussion concerning the reference frame problems.
Acknowledgements of financial support are given to
the Russian Ministry of 
Industry and Science, grant NS-1885.2003.2 (V.D.E.) and to
the National Science and Engineering
Research Council of Canada (E.L.T.).


\vfill\eject
\newpage

\begin{table}
\caption{$^3$H ground state properties with AV18+UrbIX, 
AV18+TM$\prime$ and BonnRA+TM$^\prime$ Potentials for binding
energy (EB), point charge radius (r) and probabilities of total orbital
angular momentum components in \%}
\renewcommand{\arraystretch}{1.1}
\begin{tabular}{cccc}
  \   &\  AV18 +UrbIX \  &\ \  AV18 + TM$^\prime$\ \ 
&\ \ BonnRA+TM$^\prime$\\
\hline

EB [MeV] & 8.47   & 8.47   & 8.47  \\
r [fm]   & 1.588  & 1.589  & 1.587 \\
S-wave   & 90.60  & 90.63  & 92.69 \\
P-wave   & 0.13   & 0.13   & 0.08  \\
D-wave   & 9.27   & 9.23   & 7.23  \\
\end{tabular}
\end{table}

\begin{table}
\caption{$^3$H Coulomb Sum Rule for AV18, AV18+UrbIX and MT-I/III Potentials}
\renewcommand{\arraystretch}{1.1}
\begin{tabular}{ccccc}
   $q$ [MeV/c]\  &\  $J_{max}$\  &\  AV18\  &\ \  AV18 + UrbIX\ \ &\ \ MT-I/III \\
\hline

250 & ${15\over 2}$ & 0.998 & 0.999 & 1.000  \\
300 & ${15\over 2}$ & 0.993 & 0.994 &   \\
350 & ${21\over 2}$ & 0.992 & 0.993 &   \\
400 & ${21\over 2}$ & 1.003 & 0.998 &   \\
450 & ${21\over 2}$ & 0.998 & 0.999 &   \\
500 & ${21\over 2}$ & 0.977 & 0.977 & .994 \\
\end{tabular}
\end{table}

\vfill\eject
\newpage

\begin{figure}[ht]
\includegraphics[width=6.5in]{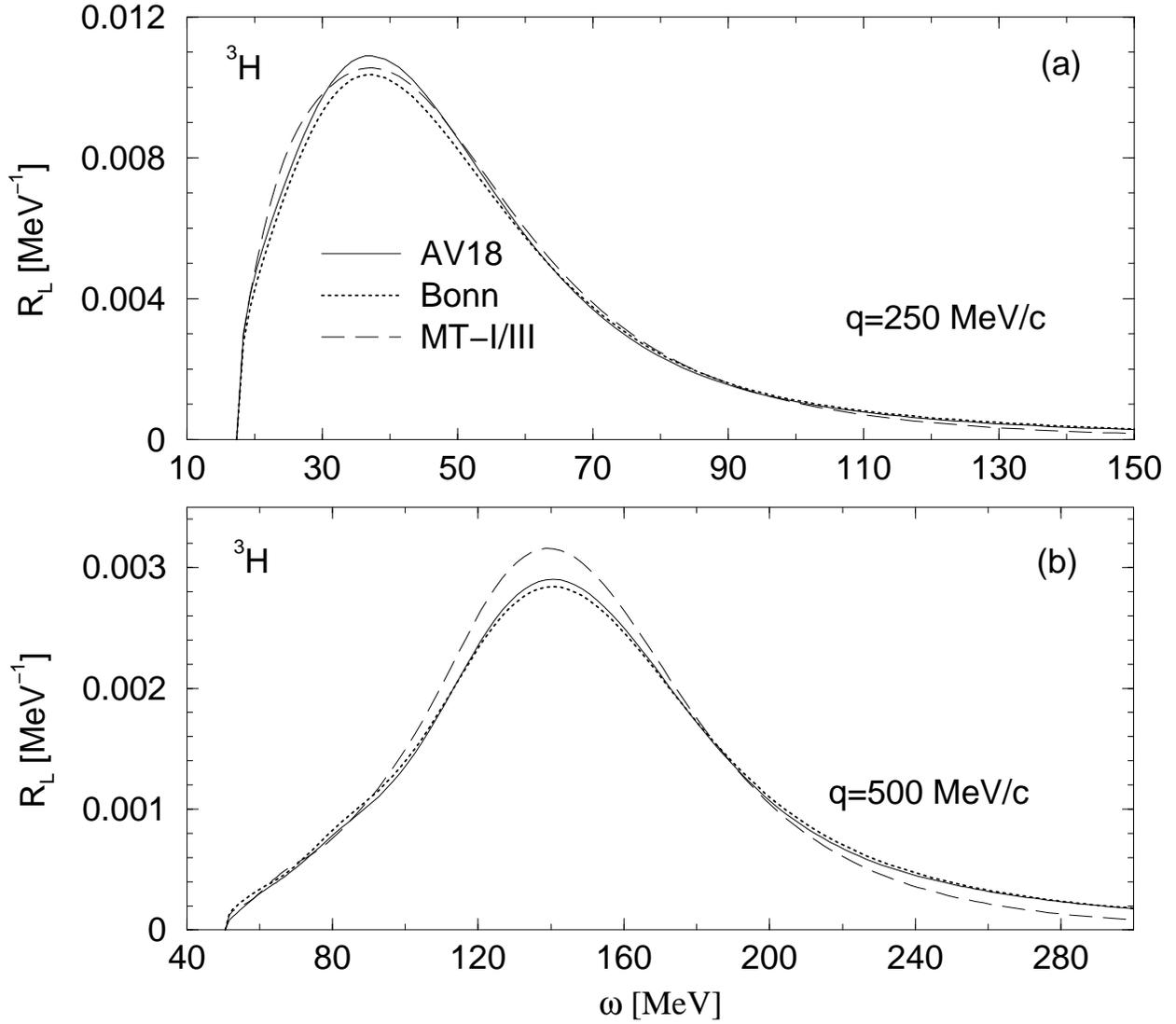}
\caption{NN potential model dependence of triton $R_L^{LAB}(q_{LAB},\omega_{LAB})$ 
at $q_{LAB}$=250 (a) and 500 (b) MeV/c (charge operator: non-relativistic plus 
DF term): AV18 (solid), BonnRA (dotted), and MT-I/III (dashed).}
\label{fig1}
\end{figure}

\begin{figure}[ht]
\includegraphics[width=6.5in]{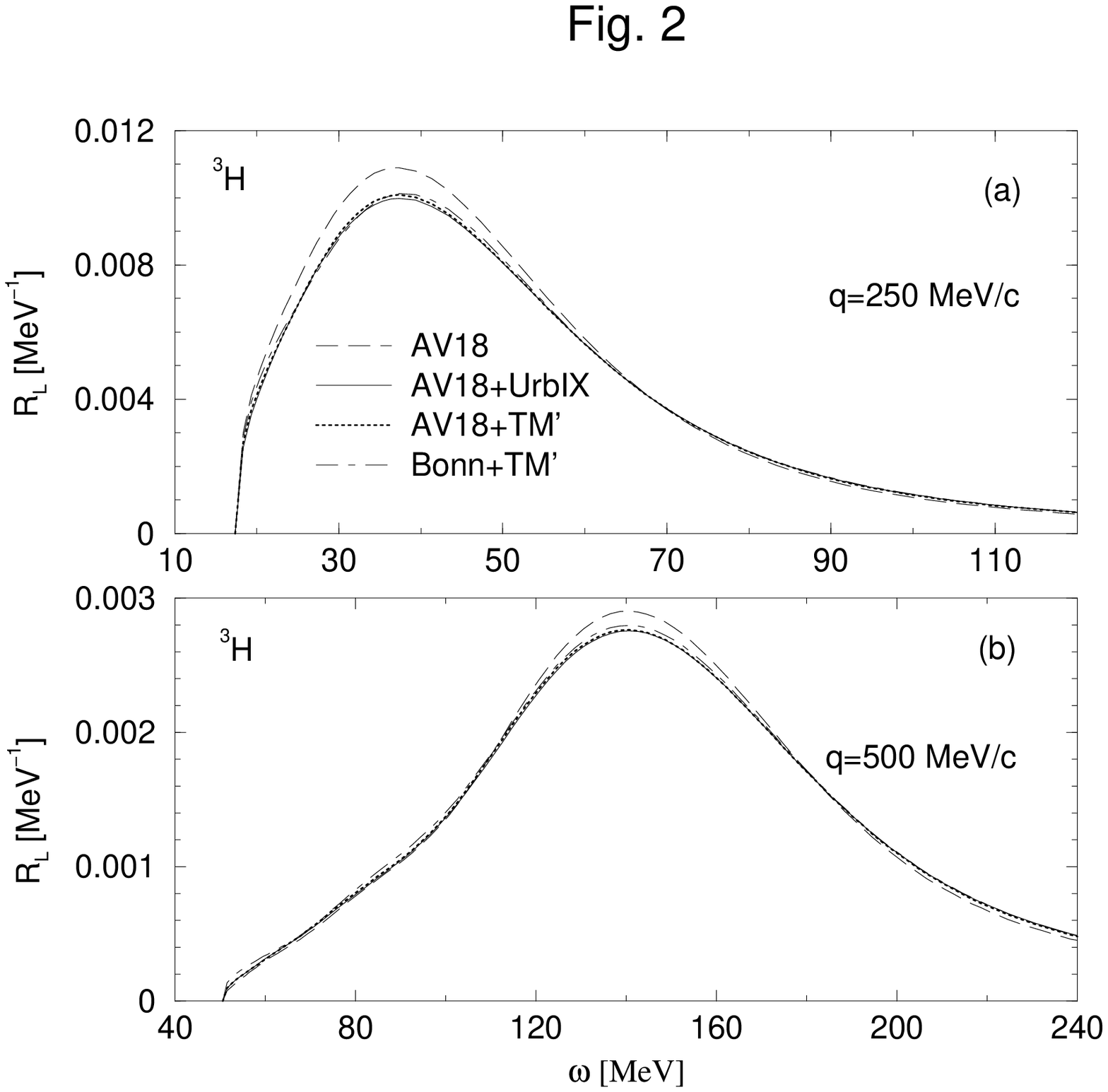}
\caption{Effect of 3N force on triton $R_L^{LAB}(q_{LAB},\omega_{LAB})$ at 
$q_{LAB}$=250 (a) and 500 (b) MeV/c (charge operator: non-relativistic plus DF 
term): AV18 (solid), AV18+UrbIX (dotted), AV18+TM$^{\prime}$ (dashed), 
and BonnRA+TM$^{\prime}$ (dash-dotted).}
\label{fig2}
\end{figure}

\begin{figure}[ht]
\includegraphics[width=6.5in]{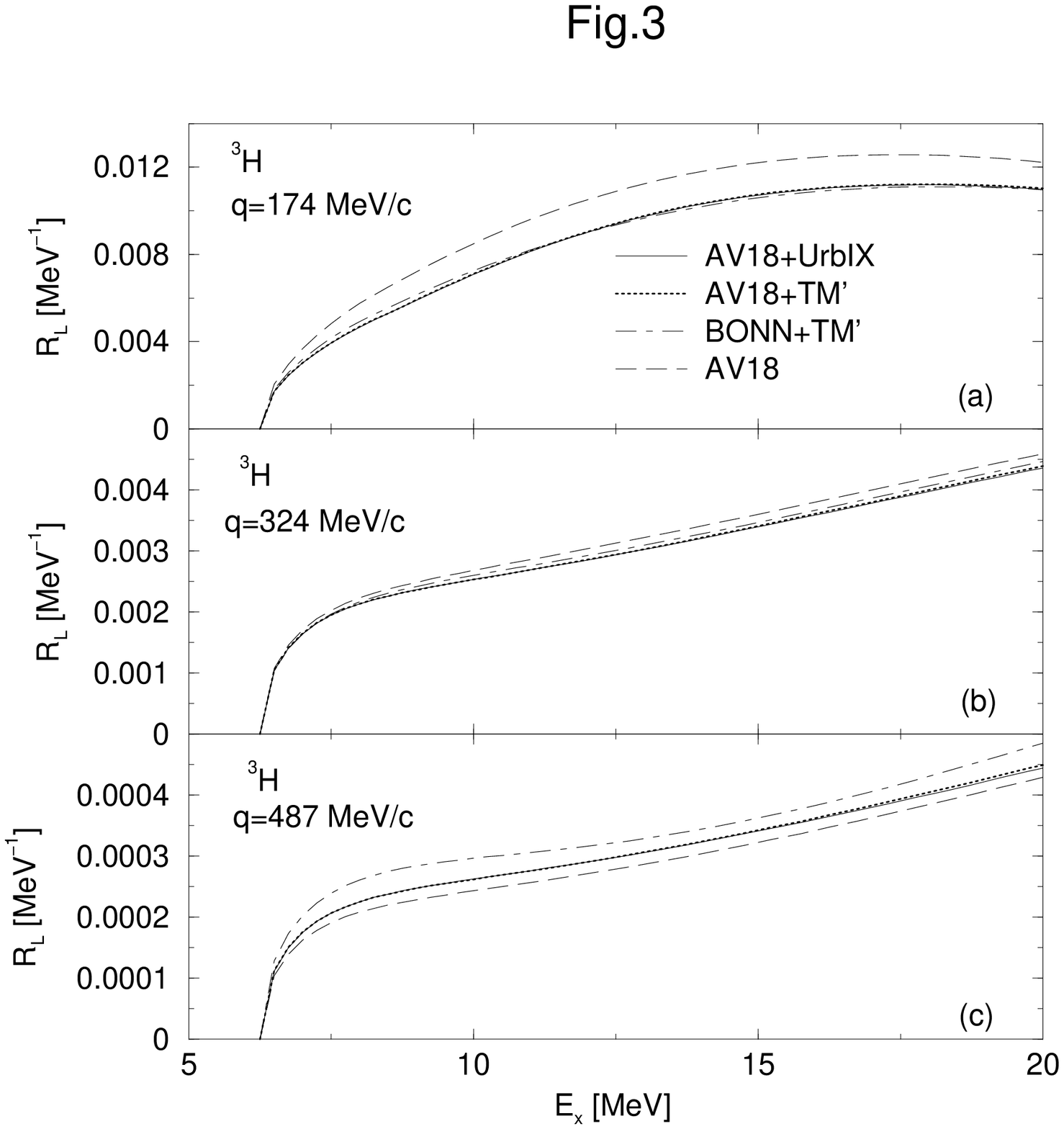}
\caption{Effect of 3N force on low-energy triton $R_L^{LAB}(q_{LAB},E_x,)$ at 
$q_{LAB}$=174 (a), 324 (b), and 487 (c) MeV/c  (charge operator:
non-relativistic plus DF term): AV18 (dashed), AV18+UrbIX (solid), 
AV18+TM$^{\prime}$ (dotted), and BonnRA+TM$^{\prime}$ (dash-dotted).}
\label{fig3}
\end{figure}

\begin{figure}[ht]
\includegraphics[width=6.5in]{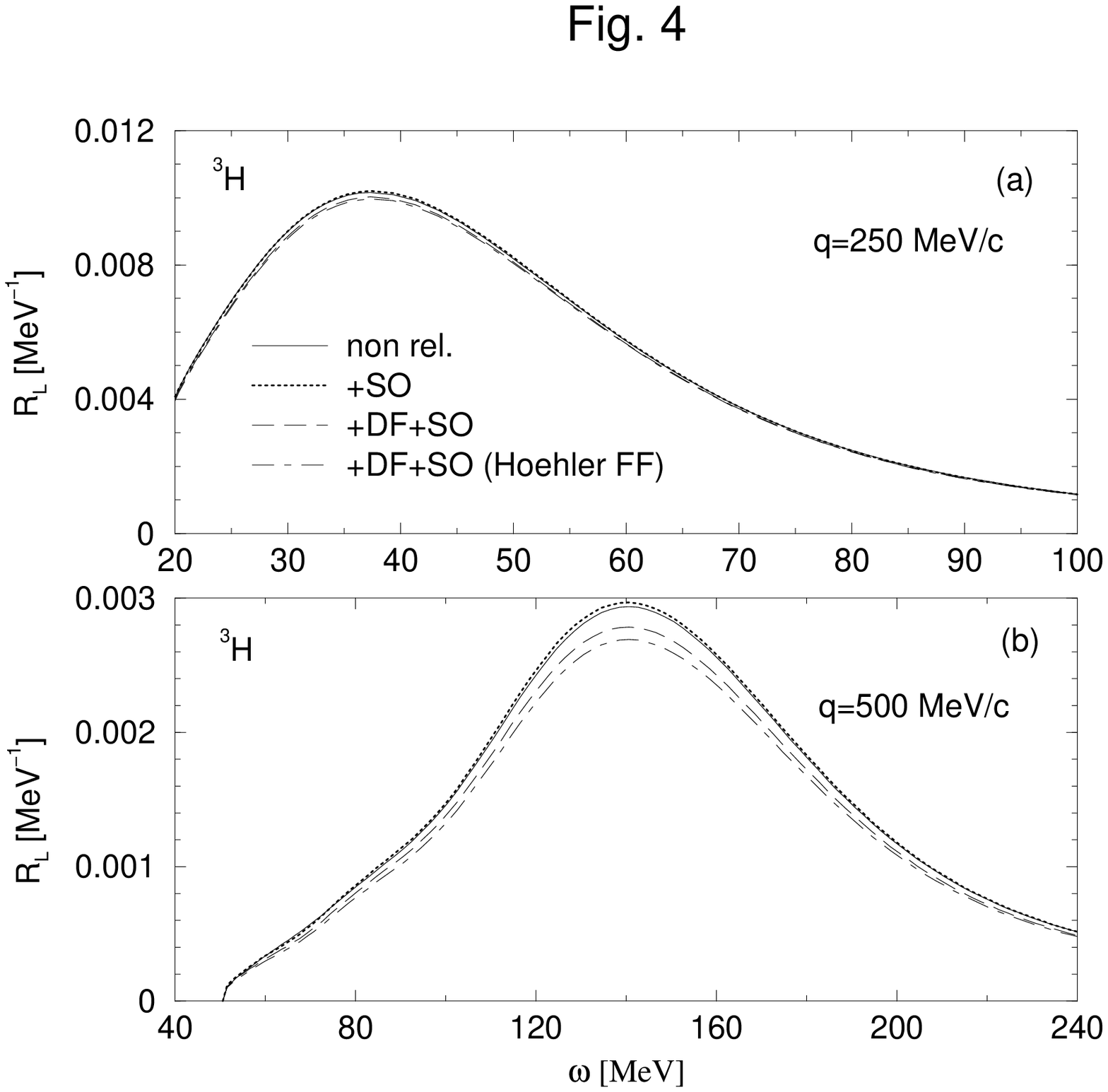}
\caption{Effect of relativistic contributions and nucleon form factor 
dependence for triton $R_L^{LAB}(q_{LAB},\omega_{LAB})$ at $q_{LAB}$=250 (a) 
and 500 (b) MeV/c (potential model: AV18+UrbIX):
non-relativistic charge operator (solid), additional inclusion of SO term 
(dotted), and total result with further inclusion of DW term (dashed); all 
three cases with neutron electric form factor from \cite{Galster} and dipole 
fit for the other three nucleon form factors. Total result also with nucleon 
form factor parametrization of \cite{Hoehler} (dash-dotted).}
\label{fig4}
\end{figure}

\begin{figure}[ht]
\includegraphics[width=6.5in]{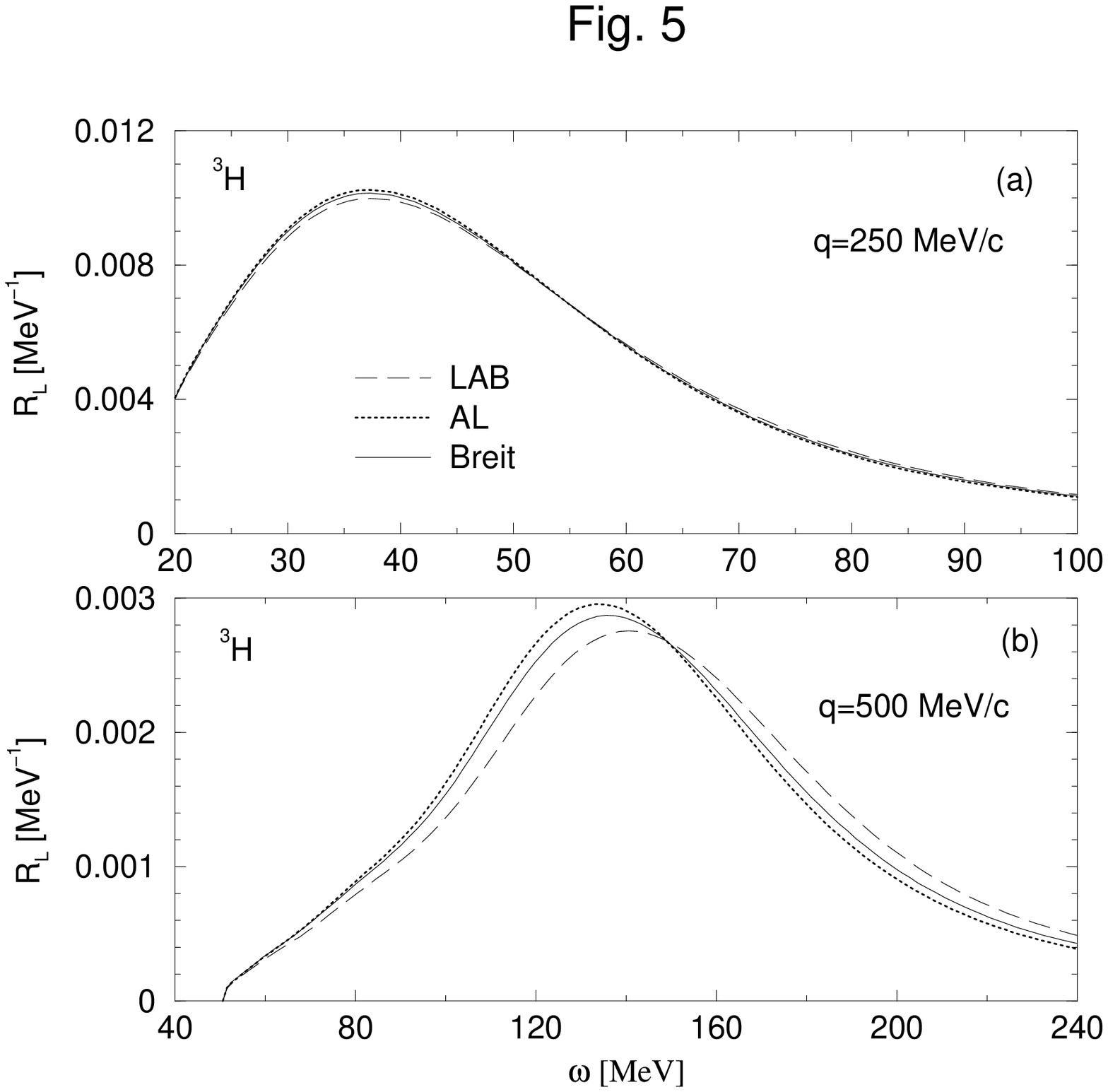}
\caption{Frame dependence of triton $R_L(q_{LAB},\omega_{LAB})$ at $q_{LAB}$=
250 (a) and 500 (b) MeV/c (potential model: AV18+UrbIX, charge operator: 
non-relativistic plus DF term): $R^{LAB}_L$ (dashed), $R^{LAB(AL)}_L$ (dotted)
and  $R^{LAB(B)}_L$ (solid).}
\label{fig5}
\end{figure}

\begin{figure}[ht]
\includegraphics[width=6.5in]{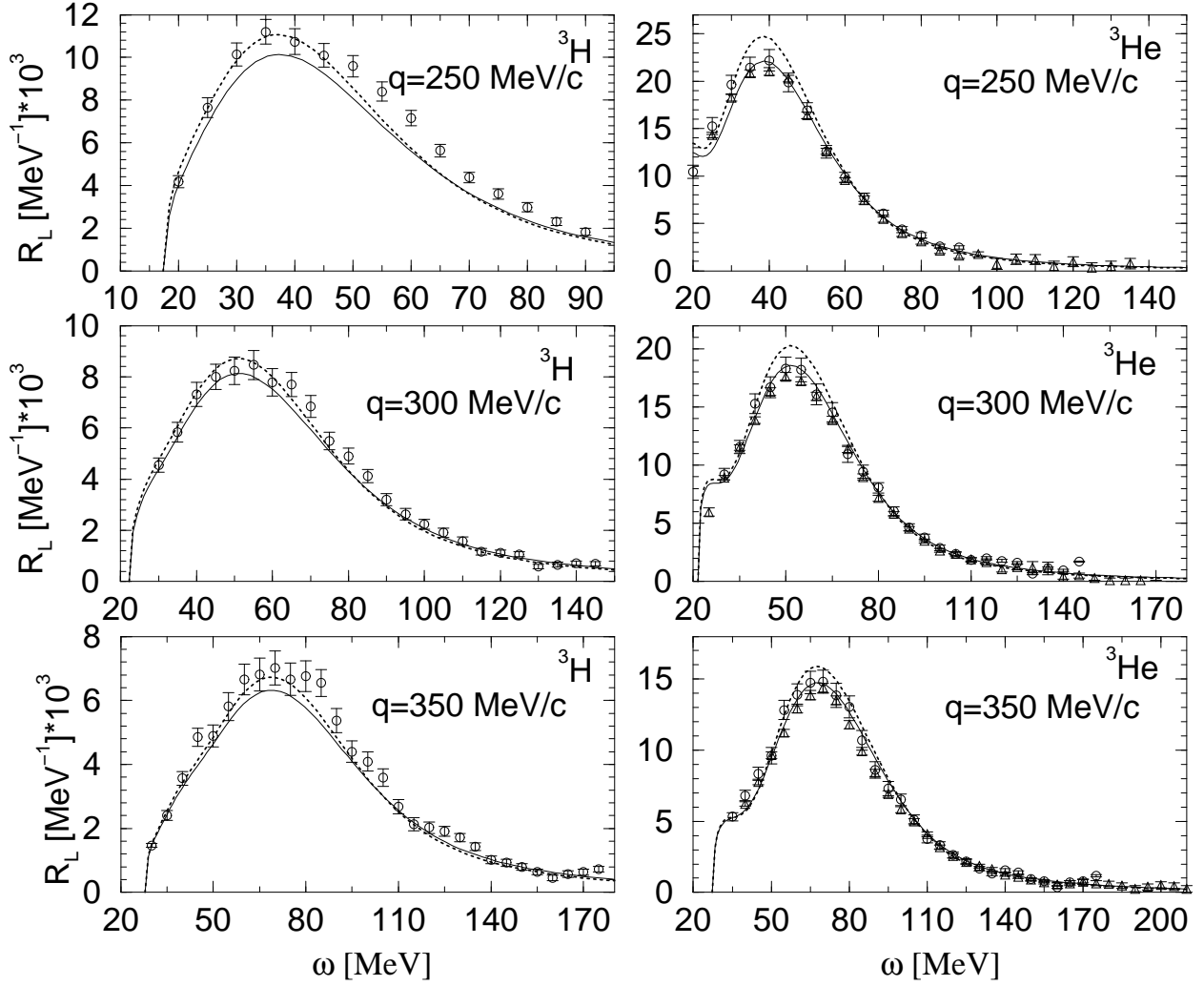}
\caption{Comparison of theoretical and experimental 
$R^{LAB(B)}_L(q_{LAB},\omega_{LAB})$ at $q_{LAB}$ as indicated in figure for 
$^3$H (left) and $^3$He (right) (charge operator: non-relativistic plus DF 
term): AV18+UrbIX potentials (solid) and AV18 potential (dotted);  
experimental data from \cite{dow} (circles) and \cite{mar,morgen} (triangles).}
\label{fig6}
\end{figure}

\begin{figure}[ht]
\includegraphics[width=6.5in]{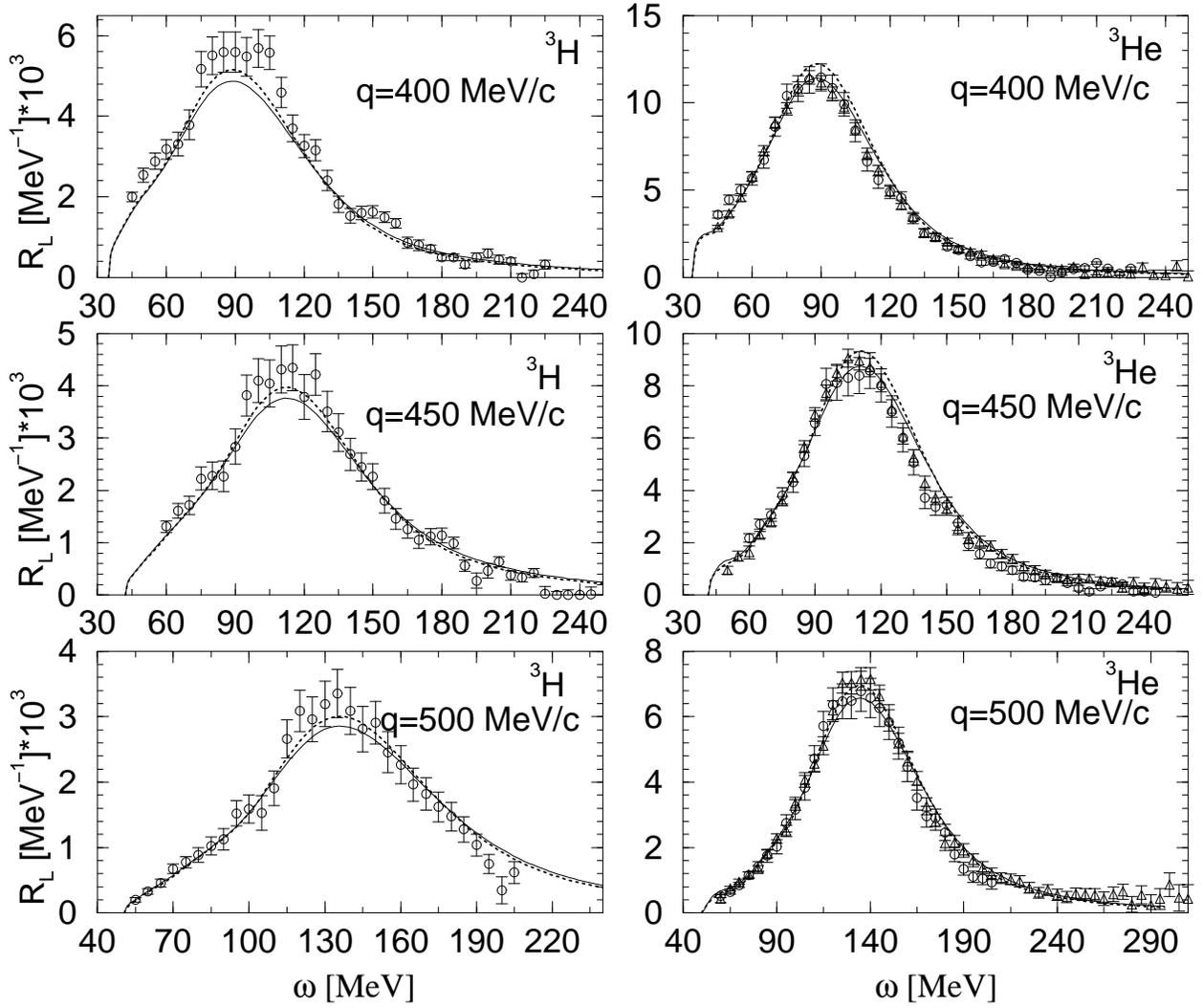}
\caption{As Fig.~6 but for different momentum transfers $q_{LAB}$ as indicated 
in figure.}
\label{fig7}
\end{figure}

\begin{figure}[ht]
\includegraphics[width=6.5in]{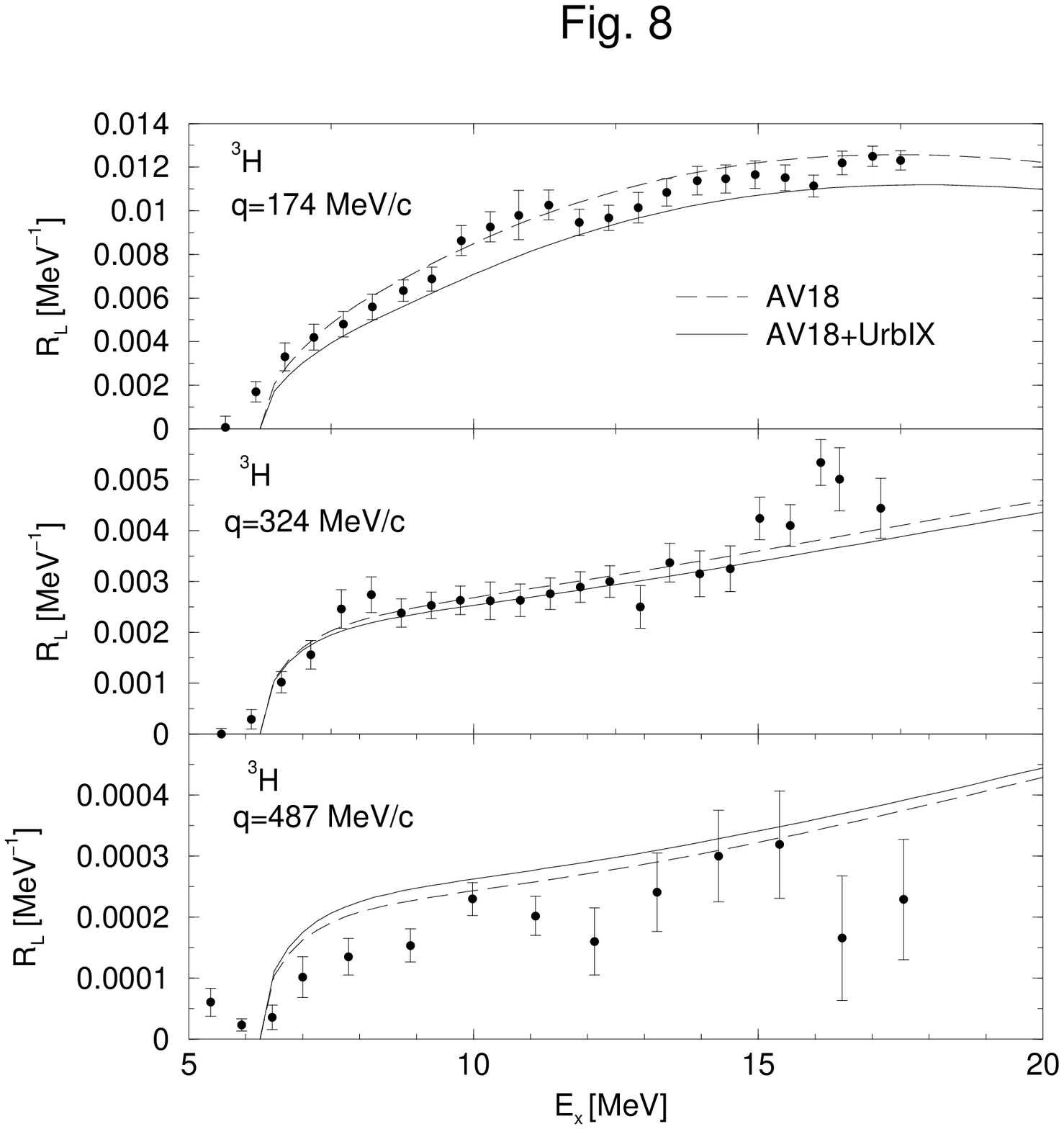}
\caption{Comparison of theoretical and experimental $R_L^{LAB}(q_{LAB},E_x)$ 
for $^3$H at $q_{LAB}$ as indicated in figure (charge operator: 
non-relativistic plus DF term): AV18+UrbIX potentials (solid) and AV18 
potential (dashed); experimental data from \cite{retz}.}
\label{fig8}
\end{figure}

\begin{figure}[ht]
\includegraphics[width=6.5in]{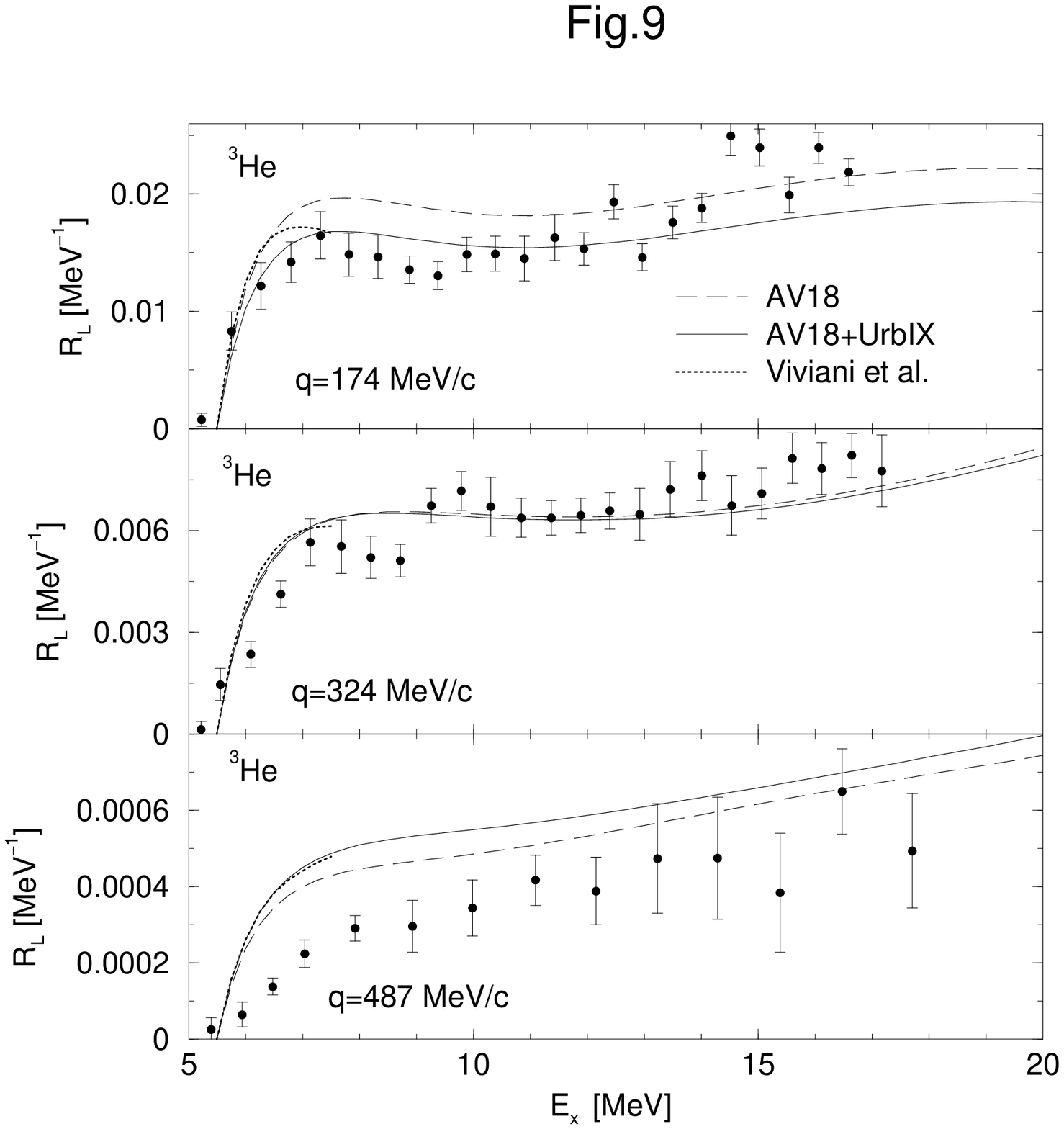}
\caption{Comparison of theoretical and experimental $R_L^{LAB}(q_{LAB},E_x)$
for $^3$He at $q_{LAB}$ as indicated in figure (charge operator: 
non-relativistic plus DF and SO terms): AV18+UrbIX potentials
(solid) and AV18 potential, but without inclusion of SO term (dashed); 
experimental data from \cite{retz}. Theoretical result from \cite{viv} (dotted)
with AV18+UrbIX potentials and same charge operators as in our AV18+UrbIX 
case.}
\label{fig9}
\end{figure}

\end{document}